# Ranking research institutions by the number of highly-cited articles per scientist[1]

*Giovanni Abramo* (corresponding author)
Laboratory for Studies of Research and Technology Transfer
Institute for System Analysis and Computer Science (IASI-CNR)
National Research Council of Italy
Via dei Taurini 19, 00185 Rome - ITALY
giovanni.abramo@uniroma2.it
Tel/fax +39 06 72597362

*Ciriaco Andrea D'Angelo*
Department of Engineering and Management
University of Rome "Tor Vergata"
Via del Politecnico 1, 00133 Rome - ITALY
dangelo@dii.uniroma2.it

**Abstract**

In the literature and on the Web we can readily find research excellence rankings for organizations and countries by either total number of *highly-cited articles (HCAs)* or by ratio of *HCA*s to total publications. Neither are indicators of efficiency. In the current work we propose an indicator of efficiency, the number of *HCA*s per scientist, which can complement the productivity indicators based on impact of total output. We apply this indicator to measure excellence in the research of Italian universities as a whole, and in each field and discipline of the hard sciences.

**Keywords**

Research evaluation; university; bibliometrics; Italy

# 1. Introduction

In Abramo & D'Angelo (2014), we provide the definition, measurement operationalization, and underlying theory of an indicator for productivity in research, named Fractional Scientific Strength (FSS). We have now used FSS over the past eight years to rank the performance of Italian professors and universities. FSS embeds both publications and citation counts, and so departs from the traditional bibliometric definitions of productivity as the number of publications per researcher. Instead, the conception of the FSS is that the more researchers publish, and are cited over a period of time, the higher is their productivity.

Productivity is the quintessential indicator of efficiency in any production system. For this, we hold that it should also be the main indicator in the assessment of performance by individual researchers and their institutions. Certainly, it cannot be the only indicator. In designing evaluation systems, the appropriate choice of performance indicators depends on the context and the policy and management objectives intended for the evaluation. The task of the bibliometrician is thus to identify and recommend the indicators most suited to the particular assessment exercise. In addition to productivity, other measures which we typically propose to policy-makers and research administrators include: the rate of concentration of unproductive researchers; the rate of concentration of top scientists (defined as authors of highly-cited publications), and the dispersion of performance within and between and research units. For all these indicators, we produce rankings that inform the decision-maker on the different quality dimensions of the individual scientists, the research units, and the institutions by field, discipline, and as a whole.

In the current work we present and apply a further indicator of performance for the research unit, in some senses complementary to the measure of research productivity (FSS). The new indicator is the number of highly-cited articles (*HCA*s) per researcher[2]. To better demonstrate the complementary character of the two indicators, we begin from the axiom that is at the basis of the productivity measures for many production systems. In the stock market, for example, the axiom would hold that the performance of two traders investing the same amount of money in two different stock portfolios bearing the same risks, is the same if the rate of return on their investments is the same. The investor can hold a portfolio of size $m$, where $m$-1 of stocks earned nothing and only one stock earned $n$ euro. The performance is considered equal to a portfolio where each of the $m$ stocks earns $n/m$ euro, all other factors constant. In the same way, with other conditions equal, a researcher publishing one publication with $n$ citations is considered to have exactly the same productivity as another researcher producing $m$ publications with $n/m$ citations each. The axiomatic concept, of a linear relationship between the scientific impact of articles and the number of their citations, could be debatable: Someone could argue that an article presenting a breakthrough discovery or radical invention, and so cited 1,000 times, is more important than 10 articles presenting

---

[2] We wish we could measure the number of HCAs per R&D spending. Unfortunately, we have no information about the resources available to each researcher, which is a common problem in most countries. Actually, we know the average cost of each researcher per academic rank. We exploit this information to reduce distortions in comparing university performance. We therefore normalize each researcher by the average salary of his/her academic rank. The actual indicator that we measure is then the number of HCAs per researcher's cost. For ease of exposition, in the following we simply refer to HCAs per researcher. This indicator should be easier to measure in those countries where the information on salaries is not available.

incremental advancements of science or technology, each one cited 100 times.

The score by the more popular performance indicators, such as all those based simply on publication counts, and the h-index would rank lower the author of one, albeit highly-cited publication. Our FSS indicator of productivity does consider such cases as indifferent. That is why, we regard as useful to flank it with another indicator that ranks research units or universities by the number of *HCA*s per researcher. Fundamentally this is still an indicator of productivity (i.e. ratio of output to input), with the difference that here the output of interest is not the overall research impact, but rather the excellent results only. Conventional wisdom would suggest to expect a positive correlation between the rankings by the two indicators at the individual level. In fact, Abramo, Cicero & D'Angelo (2014a) have shown that the most productive researchers (by FSS) are the ones that produce most of the *HCA*s.

A reasonable doubt to the reader could be whether there is any difference between the new indicator and the "concentration of top scientists", defined above as the authors of the *HCA*s. As a matter of fact, the literature suggests that these are indeed different conceptions of the measurement of the scientific excellence of institutions, as reflected in these two formulations, and that both can be usefully applied (Tijssen, 2003). The measurement can be conducted through two distinct approaches: from the perspective of the excellence of the research staff or of that of their research products. The first serves the purpose of identifying the institutions with the highest number of top scientists, regardless of the total number of top articles produced; the second is aimed to identifying the institutions that produce the highest number of top articles, regardless of whether they are produced by many scientists or only a few. The first approach is probably more appropriate for universities, where students would prefer a distribution of excellence among a number of professors in the faculty; the second approach instead might be more appropriate for research institutions, where the funding agency is concerned with maximizing the overall returns on research investments, regardless of how many scientists contribute to it. However, there could also be a dilemma for universities in considering the approaches, since they are at once educational and research institutions. We have previously adopted the approach of identifying the numbers of top scientists employed, in a study aimed at spotting the "excellent" research centers in Italy (Abramo, D'Angelo & Di Costa, 2009). There are many more examples applying the approach of identifying research institutions with the highest numbers of top articles. According to Zitt et al. (2005) *HCA*s is one of the most frequently used indicators for measurement of excellence. In the literature and on the Web we can readily find rankings of organizations and countries by either total number of *HCA*s or by ratio of *HCA*s to total publications. For example Bornmann & Leydesdorff (2011) used the ratio of 10% most-cited papers to total papers to locate the European cities producing more excellent papers than expected. Bornmann, De Moya Anegón & Leydesdorff (2012) then tested the mathematical consistency of this indicator, named "excellence rate", which is also used by SCImago in its regular *SCImago Institutions Rankings*[3]. The same indicator, but named differently "PP(10%)", is applied in the *CWTS Leiden ranking*[4] better explained in Waltman et al. (2012). This perspective in analyzing excellence has also stimulated numerous studies focused on specific sub-fields, both in the hard sciences (for example environmental sciences, Khan & Ho, 2012; or urology, Hennessey, Afshar & MacNeily, 2009) and in social sciences

---

[3] http://www.scimagoir.com/research.php, last accessed on August 31, 2015.
[4] http://www.leidenranking.com/ranking/2014, last accessed on August 31, 2015.

(for psychology, in Cho, Tse & Neely, 2012; for law, Shapiro, 1991).

Neither the absolute value of *HCA*s from the institutions nor the percentage of *HCA*s in the total of articles serve as indicators of efficiency. The first is size-dependent: other factors held equal, large organizations and countries will rank above small ones. The second is inconsistent: the percentage value of *HCA*s could decrease as the number of publications rises[5], under parity of input, and therefore this too is inappropriate to measure any efficiency dimensions of research activity.

However the number of *HCA*s per researcher is to all effects an indicator of efficiency. In fact a ranking by the number of *HCA*s per professor permits responses to the following question:

*Which individuals, research units or research organizations, working with equal production factors, produce more HCAs?*

And its corollary:

*In which field or discipline are the single institution's researchers more capable of producing HCAs, given equality of production factors?*

Many national research assessment exercises, based on the informed peer-review methodology, attempt to respond to precisely this question, producing rankings based on the best products submitted by the research institutions. However, that the objects of a national comparative evaluation should only be the best products of research, and not the entire output, appears highly debatable. In our opinion it is then more the choice of methodology that determines the evaluation objective, rather than the objective that guides the methodology. Since peer review is unable to take in the entire national scientific production, the evaluation exercises must necessarily be restricted to a subset of researchers (the Research Excellence Framework (REF) in the UK) and/or the best products (the two Italian assessment exercises, VTR and VQR, and again the REF). On the other hand, a completely bibliometric methodology would permit, at least for the hard sciences, the comparison of performance on the basis of both the totality of production, and/or of the highly-cited production alone. As well, the rankings would surely be more precise. In fact, one of the problems that afflicts research assessment by informed peer review is the inefficiency in the selection of the best products on the part of the research institutions.[6]

We should note that the indicator we present here is still subject to the usual limits intrinsic in bibliometric evaluation, implying the necessary cautions in the use of the results. In fact to measure the number of *HCA*s, bibliometricians must always refer to databases such as Web of Science (WoS) or Scopus, then ignoring those publications not included in these indexes. The methodology also ignores other forms of output, such as patents, which could in fact represent radical innovations. As partial compensation for this omission, we note that patents are often followed by publications that describe their content in the scientific arena, so the analysis of publications alone may in many cases avoid double counting. A further methodological caution is that efficiency assessments should account for all production factors, not just labor. Unfortunately the identification and calculation of value of production factors other than labor, including their share among the research fields, is a formidable task (consider for example quantification of the value of accumulated knowledge and scientific instruments, shared among university units). In most cases the bibliometricians are forced to assume that production factors other than labor are equal for all assessed units.

---

[5] Suppose one wants to compare the performance of two researchers, A and B. A, all others equal, produced one article and it is an HCA. B produced 3 articles, but only 2 are HCAs. A has a better performance than B, by the indicator "number of HCAs per total publications".

[6] For example Abramo, D'Angelo & Di Costa (2014b) estimated the error in the universities' selection of products for the hard sciences during the VQR: the scores actually achieved by the institutions are 23% to 32% worse than what could have been achieved with efficient selection of products.

After presenting the measurement methodology and the construction of the dataset in the next section, in the third section we will carry out the application of the proposed indicator to rank all Italian universities for the period 2008-2012, in each field and discipline and at the overall level. In the final section we present our conclusions.

**2. Data and methods**

To answer our research question we must first have a definition of "highly-cited" article, and choose the reference population for excellence. Furthermore, since the universities are unequal in terms of fields of research and size of staff per field, it is also necessary to address the problem of different intensities of publication across research fields (Garfield 1979; Moed et al. 1985; Butler 2007). If in one field the tendency is to publish more than in another field, then the number of *HCA*s for the researchers in the first will be greater than for those the second, all other factors equal. The measurements for comparative evaluation at the aggregate level, such as disciplines and universities, thus require particular operations in order to avoid distortion in the rankings. The identification of the threshold above which an article can be defined "highly-cited" is subjective, in general dictated by the evaluation context and the objective of the measure: it can be the top-cited 1%, 5%, 10%, etc. articles among those indexed in the same year and same field. For the current work we adopt the 10% threshold and the WoS subject category classification of articles. In the case of journals with multiple subject categories, we use the average percentile by citations in the diverse subject categories. This field classification presents few limits, especially with multidisciplinary journals, which in some cases host truly multidisciplinary articles, while in othersarticles from various fields. Alternative classification-free methods of normalization could be adopted in place of the one used here.

A further question concerns the choice of whether to use the percentile standing of the article within the world or the national population of articles indexed in WoS. The adoption of the world reference is appropriate when the aim of an evaluation is to carry out strategic analysis, for example to identify the research fields where a particular country is relatively weak or strong. With the evaluation results in hand, policy-makers may at that point choose to invest more heavily in the weak fields if these are considered strategic, not necessarily cutting back. The case is different if the assessment is aimed at comparing the efficiency of research institutions within a country. The fields researchers are involved in, should be the outcome of upstream strategic decisions. As a consequence, adopting an international reference would penalize those research groups and institutions more involved in catch-up research or active in fields where the country is not on the international frontier. The adoption of the world population is also likely to induce opportunistic behavior by research units and institutions, which would find it convenient to exit fields where the country is weak and enter those where it is strong. In the end this could negatively affect the public good. Because our aim here is to rank Italian universities, we refer to national reference, identifying the top-cited 10% of publications (for each year and subject category) within all Italian ones.[7]

Research projects frequently involve a team of researchers, as can be seen in the fact of co-authorship of publications. Productivity measures then need to account for the fractional contributions of the single units to the outputs. The contributions of the individual co-authors to the research achievement are not necessarily equal, and in some fields the authors signal the different contributions through their order in the byline. The conventions on the ordering of authors for scientific papers differ across fields (Pontille 2004; RIN 2009), thus we weight the fractional contributions of the individuals, as reflected in these conventions. Fractional contribution then equals the inverse of the number of authors, in those fields where the practice is to place the authors in simple alphabetical order, but assumes different weights in other cases. For the life sciences, widespread practice in Italy and abroad is for the authors to indicate the various contributions to the published research by the order of the names in the byline. For these disciplines, we give different weights to each co-author according to their order in the byline and the character of the co-authorship (intra-mural or extra-mural). If first and last authors belong to the same university, 40% of citations are attributed to each of them; the remaining 20% are divided among all other authors. If the first two and last two authors belong to different universities, 30% of citations are attributed to first and last authors; 15% of citations are attributed to second and last author but one; the remaining 10% are divided among all others[8]. Failure to account for the number and position of authors in the byline would result in notable ranking distortions (Abramo, D'Angelo & Rosati, 2013).

In measuring $HCAs$ per researcher, if there are differences of production factors available to each researcher, one should normalize for these. Unfortunately, relevant data are not easily available, especially at the individual level. Thus an often-necessary assumption is that the resources available to researchers within the same field are the same. A further assumption, again unless specific data are available, is that the hours devoted to research are more or less the same for each individual. Finally, as occurs for output, the value of researchers is not undifferentiated and this is reflected in the different cost of labor, which varies among research staff, both within and between units.

The productivity of full, associate and assistant professors is known to be different (Abramo, D'Angelo & Di Costa, 2011). Because the composition of research staff by academic rank varies across fields and universities, and academic rank in general determines the differentiation of salaries, to take account of costs for the production of $HCA$s we normalize each research "staff unit" by the average salary[9] of that individual's academic rank. Under the Italian university system, each academic is classified in one and only one research field, named Scientific Disciplinary Sector (SDS), of which there are 370[10], grouped in 14 disciplines, named University Disciplinary Areas (UDAs).

Considering all the above, at the field (SDS) level, the yearly average performance by $HCAs$ (which we name $P\_HCA_S$) for a university in a specific SDS $S$ is then:

---

[7] It should be noticed that a negative side effect of this choice is that researchers in weak fields may have little incentives to catch up, if the evolution of the field at international level is not monitored.
[8] The weighting values were assigned following advice from senior Italian professors in the life sciences. The values could be changed to suit different practices in other national contexts.
[9] For privacy reasons, information on individual salaries is unavailable.
[10] The complete list is accessible on http://attiministeriali.miur.it/UserFiles/115.htm, last accessed on August 31, 2015.

$$P\_HCA_S = \frac{1}{w_S} \sum_{i=1}^{N} f_i * 100$$

[1]

Where:
$w_S$ = total salary of the university research staff in *S*, in the observed period;
*N* = number of *HCAs* of the university research staff in *S*, in the period of observation;
$f_i$ = fractional contribution of researchers in the SDS of the university, to publication *i*.
The only purpose of the multiplier (100) is to make the numeric results more readable.

We can develop similar rankings at the discipline (UDA) level by aggregating the performance values of all the SDSs belonging to the UDA, normalized with respect to the national averages and weighted according to their dimension within the UDA. Thus the performance *P_HCA<sub>U</sub>* of a university in a specific UDA *U*, beginning from the performance in each SDS falling in the UDA, is:

$$P\_HCA_U = \sum_{k=1}^{N_U} \frac{P\_HCA_{S_k}}{\overline{P\_HCA_{S_k}}} \cdot \frac{w_{S_k}}{w_U}$$

[2]

With:
$w_{S_k}$ = total salary of the research staff of the university in the SDS *k*, in the observed period;
$w_U$ = total salary of the research staff of the university in the UDA *U*, in the observed period;
$N_U$ = number of SDSs of the university in the UDA *U*;
$\overline{P\_HCA_{S_k}}$ = weighted[11] average *P_HCA<sub>S</sub>* of all universities producing *HCA<sub>S</sub>*[12] in SDS k.

In analogous manner we can also arrive at the ranks of the universities as a whole.

To operationalize the above formulas, we draw on the Italian Observatory of Public Research (ORP), a database developed and maintained by the authors and derived under license from the WoS. Beginning from the raw data of Italian publications[13] indexed in WoS, we first extrapolate the top 10 percent by citations in each year and subject category. Then by applying a complex algorithm for disambiguation of the true identity of the authors and their institutional affiliations (D'Angelo, Giuffrida & Abramo, 2011), each publication is attributed to the university professors that authored it, with a harmonic average of precision and recall (F-measure) equal to 96 (error of 4%). We further reduce this error by manual disambiguation.

The period of observation of research results is 2008-2012[14]. Citations are observed on 15/05/2014[15]. Data on Italian academics in the observed period are extracted from the official database[16] maintained by the Italian Ministry of Education, Universities and Research (MIUR). The database indexes names, academic rank, affiliation, and SDS of all academics in Italian universities. At 31/12/2013 the entire Italian university population consisted of 56,600 scientists employed in the 96 universities recognized by the MIUR. About 54,000 of these were on staff for at least one year over the 2008-2012 period. It has been shown (Moed 2005) that in the so-called hard sciences, the prevalent form of codification for research output is publication in scientific journals. Thus for reasons of robustness, we examine only the nine UDAs that deal with the hard sciences,[17] and within these only those SDSs in which at least 50% of the researchers achieved at least one publication during the period observed (188 of a total 205 SDSs). Thus the dataset for the analysis includes 31,695 scientists, employed in 86 universities, authoring about 200,000 WoS publications and 21,000 *HCA*s, sorted in the UDAs as shown in Table 1.

---

[11] The weighting accounts for the relative size (in terms of cost of labor) of the SDS of each university. In other words, if the SDS of University *A* is twice as large as that of University *B*, *A*'s *P_HCA<sub>S</sub>* will weight twice as much as that of university *B*.

[12] It has been demonstrated that the average of the distribution of citations received for all cited publications of the same year and subject category is the most effective scaling factor (Abramo et al. 2012d). Because of the notable skewness of the HCAs distributions, similar to the citations distributions, we have assumed that the same occurs with HCAs. The definite choice of the most appropriate scaling factor for HCAs would require further investigation, similar to the one carried out in the above mentioned article.

[13] We exclude those document types that cannot be strictly considered as true research products, such as editorial material, meeting abstracts, replies to letters, etc.

[14] For the appropriate publication period to be observed, see Abramo, Cicero, & D'Angelo (2012b).

*Table 1: Dataset for the analysis - number of fields (SDSs), universities, research staff and WoS publications in each UDA under investigation*

| UDA | SDS | Universities | Research staff | Publications* | HCAs* |
|---|---|---|---|---|---|
| 1 - Mathematics and computer science | 9 | 72 | 2,941 | 15,982 | 1,913 |
| 2 - Physics | 8 | 65 | 2,095 | 22,160 | 2,800 |
| 3 - Chemistry | 12 | 60 | 2,781 | 25,299 | 2,961 |
| 4 - Earth sciences | 12 | 50 | 1,008 | 5,793 | 677 |
| 5 - Biology | 19 | 67 | 4,584 | 32,687 | 3,692 |
| 6 - Medicine | 49 | 65 | 9,246 | 68,504 | 7,438 |
| 7 - Agricultural and veterinary sciences | 29 | 48 | 2,598 | 13,558 | 1,266 |
| 8 - Civil engineering | 9 | 55 | 1,442 | 6,743 | 658 |
| 9 - Industrial and information engineering | 41 | 74 | 5,000 | 39,820 | 4,140 |
| Total | 188 | 86 | 31,695 | 199,811† | 21,358† |

\* *The figure refers to publications authored by at least one professor pertaining to the UDA. HCAs are at times more than 10% of the relevant publications in the UDA because they might be classified in subject categories outside the UDA.*

† *The total is less than the sum of the column data due to double counts of publications co-authored by researchers pertaining to more than one UDA.*

## 3. Rankings of universities in each field

The first level of analysis is the field or SDS. As an example we rank the universities by rate of *HCAs* in the SDS MED/18 - General surgery. For reasons of significance the comparison concerns only the universities (39 in all) that over the period 2008-2012 employed a research staff of at least two units (assistant, associate, or full professors). Table 2 presents the SDS rankings[18] according to the indicator $P\_HCA_S$ [1]. At the top of the rankings we find a university that employed only 2 professors in the SDS but with an outstanding production of *HCAs* ($P\_HCA_S$ of 13.85). The second university in the list in fact trails by almost 4 points in performance (10.11). In general the distribution of the values of performance appears significantly skewed (skewness 2.29): only 5 universities show a score over 4 points; the median is 1.04 and the average is almost double (2.11). A full 6 universities register nil performance, given that the relative research staff (a total of 26 professors for these universities) did not produce any *HCAs* over the five years under examination. The largest university has a research staff of 157 professors and shows a performance (1.21) that is just above the median. The rankings by SDS permit a strategic analysis within the individual universities. The adoption of a national reference permits the universities to identify their strong and weak SDSs in terms of what is achievable in Italy. This information can then inform the university's research strategies.

---

[15] For the citation time window that optimizes the tradeoff between accuracy of rankings and timeliness of the evaluation exercise, see Wang (2013) and Abramo, Cicero & D'Angelo (2012c).
[16] http://cercauniversita.cineca.it/php5/docenti/cerca.php, last accessed on August 31, 2015.
[17] Mathematics and computer sciences; Physics; Chemistry; Earth sciences; Biology; Medicine; Agricultural and veterinary sciences; Civil engineering; Industrial and information engineering.
[18] For privacy reasons, we hide the identity of the universities.

*Table 2: Ranking list in MED/18 - General surgery, by $P\_HCA_S$ (all Italian universities with at least 2 professors in the SDS)*

| University | Research Staff | $P\_HCA_S$ | Rank | University | Research Staff | $P\_HCA_S$ | Rank |
|---|---|---|---|---|---|---|---|
| UNIV_1 | 2 | 13.85 | 1 | UNIV_21 | 6 | 0.88 | 21 |
| UNIV_2 | 8 | 10.11 | 2 | UNIV_22 | 27 | 0.84 | 22 |
| UNIV_3 | 7 | 8.02 | 3 | UNIV_23 | 11 | 0.72 | 23 |
| UNIV_4 | 17 | 7.48 | 4 | UNIV_24 | 46 | 0.46 | 24 |
| UNIV_5 | 27 | 4.57 | 5 | UNIV_25 | 28 | 0.44 | 25 |
| UNIV_6 | 59 | 3.94 | 6 | UNIV_26 | 9 | 0.27 | 26 |
| UNIV_7 | 14 | 3.81 | 7 | UNIV_27 | 17 | 0.19 | 27 |
| UNIV_8 | 19 | 3.70 | 8 | UNIV_28 | 6 | 0.13 | 28 |
| UNIV_9 | 9 | 3.33 | 9 | UNIV_29 | 16 | 0.07 | 29 |
| UNIV_10 | 14 | 2.90 | 10 | UNIV_30 | 28 | 0.06 | 30 |
| UNIV_11 | 30 | 2.51 | 11 | UNIV_31 | 12 | 0.05 | 31 |
| UNIV_12 | 18 | 2.40 | 12 | UNIV_32 | 8 | 0.02 | 32 |
| UNIV_13 | 16 | 2.28 | 13 | UNIV_33 | 13 | 0.01 | 33 |
| UNIV_14 | 9 | 1.76 | 14 | UNIV_34 | 7 | 0 | 34 |
| UNIV_15 | 39 | 1.58 | 15 | UNIV_35 | 3 | 0 | 34 |
| UNIV_16 | 13 | 1.32 | 16 | UNIV_36 | 4 | 0 | 34 |
| UNIV_17 | 157 | 1.21 | 17 | UNIV_37 | 5 | 0 | 34 |
| UNIV_18 | 56 | 1.15 | 18 | UNIV_38 | 5 | 0 | 34 |
| UNIV_19 | 29 | 1.05 | 19 | UNIV_39 | 2 | 0 | 34 |
| UNIV_20 | 9 | 1.04 | 20 | | | | |

As an example, for UNIV_3, we can examine the performance of the SDSs in Medicine (Table 3). Overall the university has 36 Medicine SDSs that employ at least 2 professors over the five-year period examined, for a total evaluated research staff of 204 units. Eighteen of the SDSs have a performance superior to the national median. Ten place among the top 10% and five of these, MED/35 (Skin and venereal diseases), MED/13 (Endocrinology), MED/43 (Legal medicine), MED/27 (Neurosurgery) and MED/29 (Maxillofacial surgery) are actually the top national SDS. Interestingly, the absolute values of $P\_HCA$ in these 5 SDSs are significantly different (from a minimum 4.25 for MED/29 to a maximum 42.76 for MED/35), confirming the different intensity of production of *HCAs* across fields, as accounted for in the methods. Seven SDSs did not produce any *HCAs* over the period.

*Table 3: National positioning of the SDSs in Medicine at UNIV_3, by P_HCA$_S$*

| SDS | Res. Staff | P_HCA$_S$ | Rank* | Perc. | SDS | Res. Staff | P_HCA$_S$ | Rank* | Perc. |
|---|---|---|---|---|---|---|---|---|---|
| MED/35 | 4 | 42.76 | 1 of 31 | 100 | MED/38 | 9 | 4.92 | 16 of 35 | 56 |
| MED/13 | 5 | 29.64 | 1 of 34 | 100 | MED/03 | 4 | 2.30 | 14 of 29 | 54 |
| MED/43 | 5 | 8.86 | 1 of 40 | 100 | MED/01 | 6 | 4.01 | 10 of 20 | 53 |
| MED/27 | 3 | 4.89 | 1 of 22 | 100 | MED/23 | 6 | 0.34 | 12 of 22 | 48 |
| MED/29 | 3 | 4.25 | 1 of 18 | 100 | MED/11 | 3 | 3.86 | 18 of 33 | 47 |
| MED/36 | 5 | 11.58 | 2 of 38 | 97 | MED/12 | 3 | 4.19 | 16 of 29 | 46 |
| MED/25 | 6 | 11.54 | 2 of 34 | 97 | MED/40 | 4 | 1.17 | 21 of 37 | 44 |
| MED/18 | 7 | 8.02 | 3 of 39 | 95 | MED/41 | 3 | 0.13 | 23 of 36 | 37 |
| MED/08 | 9 | 10.72 | 3 of 36 | 94 | MED/42 | 5 | 0.29 | 30 of 43 | 31 |
| MED/24 | 2 | 12.26 | 3 of 30 | 93 | MED/26 | 13 | 2.84 | 29 of 40 | 28 |
| MED/07 | 5 | 2.34 | 7 of 37 | 83 | MED/28 | 12 | 0.17 | 30 of 36 | 17 |
| MED/14 | 2 | 2.15 | 6 of 23 | 77 | MED/17 | 4 | 0 | 23 of 26 | 0 |
| MED/04 | 16 | 6.78 | 11 of 44 | 77 | MED/20 | 2 | 0 | 5 of 12 | 0 |
| MED/16 | 6 | 9.30 | 7 of 24 | 74 | MED/22 | 2 | 0 | 15 of 25 | 0 |
| MED/39 | 3 | 2.73 | 7 of 21 | 70 | MED/30 | 4 | 0 | 29 of 35 | 0 |
| MED/06 | 3 | 8.08 | 10 of 24 | 61 | MED/31 | 2 | 0 | 20 of 31 | 0 |
| MED/15 | 7 | 3.88 | 12 of 29 | 61 | MED/33 | 4 | 0 | 18 of 32 | 0 |
| MED/09 | 24 | 5.17 | 17 of 41 | 60 | MED/44 | 3 | 0 | 20 of 27 | 0 |

*\* The population consists of the universities having at least 2 professors in the SDS*

## 4. Rankings of universities in each discipline

The performance of the SDSs active at the individual universities can be aggregated, with the appropriate normalization and weighting, to obtain the university performance at the level of the disciplines (UDA). Table 4 presents the example of the rankings resulting from the application of this procedure [2] to the Italian universities active in Physics, composed of 8 SDSs. For reasons of significance the evaluation concerns only those universities (44 in all) that employed a research staff of at least 10 units in the UDA over the period 2008-2012.

Only two universities register a value of $P\_HCA_U$ greater than 2, and none have a nil performance. In general the distribution of performance appears less skewed than that at the SDS level. The skewness is 1.39 and the average and the median differ little (1.02 and 0.94). It is important to observe that in comparing the ranks by size of university and by performance, there is a very weak correlation: The Spearman correlation index results as 0.09. The smallest university (10 professors in the UDA) places 26[th] in the ranking. The two universities with 11 professors are at 3[rd] and 43[rd] place. The three universities with more than 100 professors in the UDA rank 7[th], 9[th] and 21[st]. The very weak correlation confirms previous studies of research activities that demonstrate constant returns to scale (Abramo, Cicero & D'Angelo, 2012a; Bonaccorsi & Daraio 2005; Seglen & Aksnes, 2000; Golden & Carstensen, 1992) and scope (Abramo, D'Angelo & Di Costa, 2014c), here again showing constant returns to scale for the production of *HCA*s.

*Table 4: Ranking list of Italian universities active in Physics, by P_HCA (considering only universities with at least 10 professors in the UDA)*

| University | Research Staff | P_HCA | Rank | University | Research Staff | P_HCA | Rank |
|---|---|---|---|---|---|---|---|
| UNIV_40 | 59 | 2.80 | 1 | UNIV_19 | 129 | 0.94 | 21 |
| UNIV_41 | 12 | 2.20 | 2 | UNIV_29 | 74 | 0.89 | 24 |
| UNIV_27 | 11 | 1.83 | 3 | UNIV_48 | 64 | 0.85 | 25 |
| UNIV_42 | 15 | 1.78 | 4 | UNIV_49 | 10 | 0.84 | 26 |
| UNIV_43 | 32 | 1.62 | 5 | UNIV_37 | 44 | 0.83 | 27 |
| UNIV_23 | 14 | 1.42 | 6 | UNIV_50 | 46 | 0.82 | 28 |
| UNIV_17 | 132 | 1.39 | 7 | UNIV_20 | 17 | 0.80 | 29 |
| UNIV_7 | 78 | 1.38 | 8 | UNIV_12 | 56 | 0.77 | 30 |
| UNIV_5 | 113 | 1.36 | 9 | UNIV_9 | 31 | 0.76 | 31 |
| UNIV_11 | 75 | 1.29 | 10 | UNIV_18 | 71 | 0.75 | 32 |
| UNIV_44 | 35 | 1.28 | 11 | UNIV_39 | 39 | 0.67 | 33 |
| UNIV_15 | 18 | 1.26 | 12 | UNIV_2 | 14 | 0.61 | 34 |
| UNIV_8 | 71 | 1.26 | 12 | UNIV_25 | 86 | 0.61 | 34 |
| UNIV_21 | 54 | 1.23 | 14 | UNIV_14 | 21 | 0.60 | 36 |
| UNIV_45 | 47 | 1.12 | 15 | UNIV_30 | 53 | 0.60 | 36 |
| UNIV_10 | 29 | 1.07 | 16 | UNIV_32 | 31 | 0.57 | 38 |
| UNIV_46 | 17 | 1.05 | 17 | UNIV_33 | 38 | 0.56 | 39 |
| UNIV_22 | 41 | 1.04 | 18 | UNIV_6 | 14 | 0.53 | 40 |
| UNIV_35 | 14 | 1.00 | 19 | UNIV_16 | 37 | 0.44 | 41 |
| UNIV_4 | 93 | 0.96 | 20 | UNIV_31 | 40 | 0.42 | 42 |
| UNIV_47 | 37 | 0.94 | 21 | UNIV_13 | 11 | 0.35 | 43 |
| UNIV_28 | 56 | 0.94 | 21 | UNIV_24 | 44 | 0.26 | 44 |

Similar to the analysis at the SDS level we now return to the perspective of the research administrator, to compare the performance of the UDAs in a single university. Table 5 presents the comparative evaluation of performance for the case of UNIV_3, in the UDAs (5 in all) where the institution employs more than 10 professors. In UDA 6 (Medicine), which is the most important in size (209 professors out of 352 total evaluated faculty), the university's performance is in the top 10% at national level. In other cases (UDA 1, Mathematics; UDA 5, Biology; UDA 7, Agriculture and veterinary science), the university is still in the national top 20%. However the university's performance in Industrial and information engineering appears very limited, at 41st out of 51 universities. Still, this is the smallest of the university's UDAs (only 12 professors), and at this scale it remains indecisive in the overall performance. The overall performance is in fact seen in the last line of Table 5, obtained by extending summation [2] to all the university's SDSs: the institution places exceptionally well at the national level, at 5th out of 63 universities.

*Table 5: Rank by $P\_HCA_U$ of the UDAs at UNIV_3*

| UDA | Research Staff | $P\_HCA_U$ | Rank* | Percentile |
|---|---|---|---|---|
| 1 | 35 | 1.25 | 8 of 50 | 86 |
| 5 | 64 | 1.00 | 10 of 52 | 82 |
| 6 | 209 | 1.41 | 5 of 44 | 91 |
| 7 | 17 | 0.99 | 6 of 29 | 82 |
| 9 | 12 | 0.48 | 41 of 51 | 20 |
| Total | 352 | 1.28 | 5 of 63 | 94 |

*\* The population consists of the universities having at least 10 professors in the UDA*

In Table 6 we provide the complete ranking list of the universities by overall P_HCA. For reasons of significance, we consider only the universities with at least 30 units of research staff in the SDSs considered. The first 4 universities in the list show outstanding performances: in effect their removal would reduce the skewness of the distribution from 2.53 to 0.23 and give perfect superimposition of the mean and median.

To assess the impact of salary normalization on the rankings, we repeated the analysis without normalizing by the salaries of professors. Practically, in formula [1] we substituted $w_S$ with the number of years of work in the period under observation. Table 7 shows the differences between the two rankings. The correlation of the two ranking lists are very high in all UDAs: Spearman ρ is never below 0.97. However, a few shifts in rank are noticeable in some UDAs. The highest average shift (2.3 positions) concerns Industrial and information engineering (UDA 9), a discipline where 80.4% of the 51 universities with at least 10 units of research staff, change position in rank. Also in Biology (UDA 5) few shifts are not negligible, affecting 73% of the 52 universities assessed, with an average of 2.1 positions and a maximum shift of 11 positions by a university. falling from the 25[th] position in the ranking to 36[th].

*Table 6: Ranking list of Italian universities, on the basis of P_HCA*

| University | Research Staff | P_HCA | Rank | University | Research Staff | P_HCA | Rank |
|---|---|---|---|---|---|---|---|
| UNIV_1 | 63 | 2.78 | 1 | UNIV_56 | 138 | 0.74 | 33 |
| UNIV_42 | 40 | 2.33 | 2 | UNIV_20 | 385 | 0.74 | 34 |
| UNIV_51 | 55 | 2.14 | 3 | UNIV_13 | 439 | 0.73 | 35 |
| UNIV_43 | 57 | 1.61 | 4 | UNIV_37 | 377 | 0.73 | 36 |
| UNIV_3 | 352 | 1.28 | 5 | UNIV_25 | 924 | 0.73 | 37 |
| UNIV_5 | 1,338 | 1.20 | 6 | UNIV_9 | 535 | 0.72 | 38 |
| UNIV_34 | 98 | 1.19 | 7 | UNIV_12 | 823 | 0.71 | 39 |
| UNIV_11 | 1,387 | 1.11 | 8 | UNIV_38 | 168 | 0.70 | 40 |
| UNIV_44 | 223 | 1.09 | 9 | UNIV_19 | 1,646 | 0.70 | 41 |
| UNIV_8 | 1,098 | 1.09 | 10 | UNIV_22 | 861 | 0.68 | 42 |
| UNIV_40 | 914 | 1.01 | 11 | UNIV_57 | 137 | 0.67 | 43 |
| UNIV_6 | 703 | 0.99 | 12 | UNIV_17 | 2,376 | 0.63 | 44 |
| UNIV_4 | 1,518 | 0.98 | 13 | UNIV_18 | 845 | 0.62 | 45 |
| UNIV_29 | 1,000 | 0.95 | 14 | UNIV_58 | 124 | 0.58 | 46 |
| UNIV_52 | 49 | 0.94 | 15 | UNIV_59 | 88 | 0.56 | 47 |
| UNIV_28 | 413 | 0.93 | 16 | UNIV_15 | 608 | 0.56 | 48 |
| UNIV_53 | 159 | 0.92 | 17 | UNIV_41 | 229 | 0.56 | 49 |
| UNIV_39 | 452 | 0.88 | 18 | UNIV_60 | 326 | 0.55 | 50 |
| UNIV_45 | 453 | 0.86 | 19 | UNIV_36 | 153 | 0.52 | 51 |
| UNIV_48 | 260 | 0.85 | 20 | UNIV_61 | 115 | 0.52 | 52 |
| UNIV_10 | 696 | 0.83 | 21 | UNIV_30 | 952 | 0.51 | 53 |
| UNIV_23 | 396 | 0.83 | 22 | UNIV_62 | 107 | 0.48 | 54 |
| UNIV_47 | 637 | 0.81 | 23 | UNIV_16 | 575 | 0.47 | 55 |
| UNIV_54 | 117 | 0.80 | 24 | UNIV_24 | 783 | 0.45 | 56 |
| UNIV_35 | 200 | 0.80 | 25 | UNIV_63 | 427 | 0.44 | 57 |
| UNIV_50 | 282 | 0.80 | 26 | UNIV_46 | 206 | 0.41 | 58 |
| UNIV_32 | 408 | 0.79 | 27 | UNIV_26 | 391 | 0.39 | 59 |
| UNIV_2 | 398 | 0.79 | 28 | UNIV_64 | 33 | 0.39 | 60 |
| UNIV_21 | 637 | 0.78 | 29 | UNIV_49 | 235 | 0.37 | 61 |
| UNIV_55 | 258 | 0.77 | 30 | UNIV_65 | 80 | 0.36 | 62 |
| UNIV_7 | 1,022 | 0.77 | 31 | UNIV_66 | 133 | 0.29 | 63 |
| UNIV_33 | 618 | 0.76 | 32 | | | | |

*Table 7: Comparison of ranking lists of Italian universities, on the basis of P_HCA with and without salary normalization*

| UDA | Universities | Spearman ρ | Shifintg in rank | Average shift | Max shift | Average percentile shift | Max percentile shift |
|---|---|---|---|---|---|---|---|
| 1 | 50 | 0.990 | 76.0% | 1.5 | 5 | 3.0 | 10.2 |
| 2 | 44 | 0.986 | 75.0% | 1.6 | 6 | 3.7 | 14.0 |
| 3 | 43 | 0.976 | 67.4% | 1.9 | 8 | 4.5 | 19.0 |
| 4 | 29 | 0.991 | 51.7% | 0.8 | 3 | 2.7 | 10.7 |
| 5 | 52 | 0.980 | 73.1% | 2.1 | 11 | 4.1 | 21.6 |
| 6 | 44 | 0.992 | 56.8% | 1.0 | 7 | 2.3 | 16.3 |
| 7 | 29 | 0.990 | 58.6% | 0.8 | 3 | 3.0 | 10.7 |
| 8 | 36 | 0.989 | 50.0% | 0.9 | 5 | 2.7 | 14.3 |
| 9 | 51 | 0.975 | 80.4% | 2.3 | 9 | 4.6 | 18.0 |
| Total | 63 | 0.988 | 69.8% | 1.8 | 9 | 3.0 | 14.5 |

## 5. Conclusions

Productivity is the quintessential indicator of efficiency in any production system. For this, we hold that it must be the principle indicator to assess the performance of individuals and research institutions. In some contexts and for specific policy and management objectives, it may be appropriate to integrate the measure of productivity based on overall articles with that based on excellent results only. For this purpose, the current paper has provided an additional indicator: the number of *HCAs* per professor. Through this indicator it is possible to identify which organizations, under parity in labor force, produce more (or less) highly cited results. Also, within each research institution, it is possible to identify which fields or disciplines produce more (or less) highly cited results.

The assumptions and limits of the proposed indicator are the same as those of FSS and any efficiency indicators in general. The ratio of output to input should account for all research results (excellent results in this case) and all production factors. HCAs indexed in WoS do not necessarily include all top publications (as indexed in Scopus for example), and do not include for sure patents and other types of codification of new knowledge. The production factors other than labor, and the time devoted by single researchers to research are generally unknown. Performance rankings are then affected by a degree of uncertainty that decreases with the amount of information embedded in the measures. Anyway, we have noticed a substantial convergence of the evaluation outcomes, with and without normalizing by the staff salaries.

All that said, there is a value added in the proposed indicator. We note in fact that those national research exercises based on (informed) peer-review methodologies, such as the REF in the U.K. or the VQR in Italy, attempt to measure excellence in research. The rankings from these evaluations are necessarily based on only the few best products as submitted by the research institutions, and are inflicted by inefficiencies in the selection of these products, yet the exercises still demonstrate very high costs and times. The use of the proposed bibliometric indicator, could help avoid the above said inefficiencies, while not adding more limits.

The application of the indicator to the evaluation of the Italian university system at various levels (field, discipline and overall institutions) provides a useful test for the empirical evaluation of the indicator itself, and of the method of its calculation at the different levels. Further extensions of the research could include the analysis of the correlation between rankings by *P_HCAs* with those by the concentration of top scientists, as well as those by research productivity (*FSS*). A further useful in-depth analysis could be to compare the rankings by *P_HCAs* to the university rankings from the VQR, the national informed peer review exercise conducted in Italy for the period 2004-2010.

**References**

Abramo, G., D'Angelo, C.A. & Di Costa, F. (2009). Mapping excellence in national research systems: the case of Italy. *Evaluation Review*, 33(2), 159-188.
Abramo, G., D'Angelo, C.A. & Di Costa, F. (2011). Research productivity: are higher academic ranks more productive than lower ones? *Scientometrics*, 88(3), 915-928.
Abramo, G., Cicero, T. & D'Angelo, C.A. (2012a). Revisiting size effects in higher education research productivity. *Higher Education*, 63(6), 701-717.
Abramo, G., Cicero, T. & D'Angelo, C.A. (2012b). What is the appropriate length of the publication period over which to assess research performance? *Scientometrics*, 93(3), 1005-1017.
Abramo, G., Cicero, T. & D'Angelo, C.A. (2012c). A sensitivity analysis of researchers' productivity rankings to the time of citation observation. *Journal of Informetrics*, 6(2), 192–201.
Abramo, G., Cicero, T. & D'Angelo, C.A. (2012d). Revisiting the scaling of citations for research assessment. *Journal of Informetrics*, 6(4), 470–479.
Abramo, G., D'Angelo, C.A. & Rosati, F. (2013). The importance of accounting for the number of co-authors and their order when assessing research performance at the individual level in the life sciences. *Journal of Informetrics,* 7(1), 198–208.
Abramo, G., Cicero T. & D'Angelo, C.A. (2014a). Are the authors of highly-cited articles also the most productive ones? *Journal of Informetrics*, 8(1), 89-97.
Abramo, G., D'Angelo, C.A. & Di Costa, F. (2014b). Inefficiency in selecting products for submission to national research assessment exercises. *Scientometrics,* 98(3), 2069-2086.
Abramo, G., D'Angelo, C.A. & Di Costa, F. (2014c). Investigating returns to scope of research fields in universities. *Higher Education*, 68(1), 69-85.
Abramo, G. & D'Angelo, C.A. (2014). How do you define and measure research productivity? *Scientometrics,* 101(2), 1129-1144.
Bonaccorsi, A. & Daraio, C. (2005). Exploring size and agglomeration effects on public research productivity. *Scientometrics*, 63(1), 87-120.
Bornmann, L. & Leydesdorff, L. (2011). Which cities produce more excellent papers than can be expected? A new mapping approach, using Google Maps, based on statistical significance testing. *Journal of the American Society for Information Science and Technology*, 62(10), 1954-1962.
Bornmann, L., De Moya Anegón, F. & Leydesdorff, L. (2012) The new Excellence Indicator in the World Report of the SCImago Institutions Rankings 2011. *Journal of Informetrics*, 6(2), 333-335.
Butler, L. (2007). Assessing university research: A plea for a balanced approach. *Science and Public Policy*, 34(8), 565-574.

Cho, K.W., Tse, C.S. & Neely, J.H. (2012). Citation rates for experimental psychology articles published between 1950 and 2004: Top-cited articles in behavioral cognitive psychology. *Memory & cognition*, 40(7), 1132-1161.

D'Angelo, C.A., Giuffrida, C. & Abramo, G. (2011). A heuristic approach to author name disambiguation in large-scale bibliometric databases. *Journal of the American Society for Information Science and Technology*, 62(2), 257–269.

Garfield, E. (1979). Is citation analysis a legitimate evaluation tool? *Scientometrics,* 1(4), 359-375.

Golden, J. & Carstensen, F.V. (1992). Academic research productivity, department size and organization: Further results, comment. *Economics of Education Review,* 11(2), 169-171.

Hennessey, K., Afshar, K. & MacNeily, A.E. (2009). The top 100 cited articles in urology. *Canadian Urological Association Journal*, 3(4), 293.

Khan, M.A. & Ho, Y.S. (2012). Top-cited articles in environmental sciences: Merits and demerits of citation analysis. *Science of the Total Environment*, 431, 122-127.

Moed H.F., Buger W.J., Franfort J.G., & van Raan A.F.J., (1985). The use of bibliometric data for the measurement of university research performance. *Research Policy*, 14(3), 131-149.

Moed, H.F. (2005). *Citation Analysis in Research Evaluation*. Springer, ISBN: 978-1-4020-3713-9

Pontille, D. (2004). *La Signature Scientifique: Une Sociologie Pragmatique de l'Attribution*. CNRS ÉDITIONS, Paris, 2004.

RIN (Research Information Network) (2009). *Communicating Knowledge: How and Why Researchers Publish and Disseminate Their Findings*. London, UK: RIN. Retrieved August 31, 2015 from: http://www.rin.ac.uk/our-work/communicating-and-disseminating-research/communicating-knowledge-how-and-why-researchers-pu.

Seglen, P.O. & Asknes, D.G. (2000). Scientific productivity and group size: A bibliometric analysis of Norwegian microbiological research. *Scientometrics,* 49(1), 123-143.

Shapiro, F.R. (1991). The most-cited articles from the Yale Law Journal. *Yale Law Journal*, 100(5), 1449-1514.

Tijssen, R.J.W. (2003). Scoreboards of research excellence. *Research Evaluation*, 12(2), 91–103.

Waltman, L., Calero-Medina, C., Kosten, J., Noyons, E.C.M., Tijssen, R.J.W., Van Eck, N.J., Van Leeuwen, T.N., Van Raan, A.F.J., Visser, M.S. & Wouters, P. (2012). The Leiden Ranking 2011/2012: Data collection, indicators, and interpretation. *Journal of the American Society for Information Science and Technology*, 63(12), 2419-2432.

Wang, J (2013). Citation time window choice for research impact evaluation. *Scientometrics*, 94(3), 851-872.

Zitt, M., Ramanana-Rahary, S., Bassecoulard, E. (2005). Relativity of Citation Performance and Excellence Measures: From Cross-Field to Cross-Scale Effects of Field-Normalisation. *Scientometrics*, 63(2), 373-401.